\documentclass[acmsmall,screen,nonacm]{acmart}

\acmJournal{PACMPL}
\acmVolume{1}
\acmNumber{OOPSLA}
\acmArticle{1}
\acmYear{2024}
\acmMonth{4}
\acmDOI{} %
\startPage{1}

\setcopyright{none} %

\usepackage{cleveref}
\usepackage[english=british]{csquotes}	%

\newcommand\basedir{./}
\usepackage{\basedir iris}
\usepackage{\basedir rusthornbelt}

\usepackage{relsize}    %
\usepackage{mathpartir}
\usepackage{makecell}   %
\usepackage{tikz}
\usetikzlibrary{tikzmark}
\usetikzlibrary{arrows.meta}
\usetikzlibrary{graphs}
\usepackage[most]{tcolorbox}
\usepackage{listings}
\usepackage{adjustbox}   %

\definecolor{White}{gray}{1}
\definecolor{DirtyWhite}{gray}{0.96}
\definecolor{LightestGray}{gray}{0.95}
\definecolor{LightGray}{gray}{0.89}
\definecolor{Gray}{gray}{0.60}
\definecolor{DarkGray}{gray}{0.25}
\definecolor{DarkGreen}{RGB}{0,90,0} %
\definecolor{DarkBlue}{RGB}{30,30,150}
\definecolor{DarkRed}{RGB}{180,20,20} %
\definecolor{LightBrown}{RGB}{225,225,219}
\definecolor{DarkBrown}{RGB}{59,46,42}
\definecolor{Orange}{RGB}{200,200,20}
\definecolor{Goldenrod}{RGB}{255,223,66}
\colorlet{LightYellow}{Goldenrod!45!white}

\colorlet{CodeKeyword}{DarkBlue}  %
\colorlet{CodeSpecial}{DarkGreen} %
\colorlet{CodeString}{DarkRed}    %
\colorlet{CodeComment}{DarkGray}  %
\colorlet{CodeBack}{LightestGray}
\colorlet{CodeFrame}{Gray}
\colorlet{CodeInlineBack}{DirtyWhite}
\colorlet{CodeInlineFrame}{LightGray}
\colorlet{OutputBack}{LightestGray}
\colorlet{OutputFrame}{Gray}
\colorlet{KeyProblemBack}{LightYellow}
\colorlet{KeyProblemFrame}{black}

\lstdefinelanguage{Viper}
{
	sensitive,
	morekeywords=[1]{
		field, domain, axiom, function, predicate, method,
		returns, requires, ensures, invariant, decreases,
		old, acc, none, write, var, let, in, forall, exists, unfolding,
		assert, assume, inhale, package, apply, exhale, havoc, fold, unfold,
		label, goto, if, else, while,
	},
	morekeywords=[2]{
	},
	morekeywords=[3]{
	},
	morekeywords=[4]{
		Int, Bool, Ref, Perm,
	},
	morekeywords=[5]{
	},
	morekeywords=[6]{
		true, false
	},
	morecomment=[l]{//},
	morecomment=[s]{/*}{*/},
}

\lstdefinestyle{colouredViper}
{
	basicstyle=,
	identifierstyle=,
	keywordstyle=[1]\bfseries\color{CodeKeyword},  %
	keywordstyle=[2]\color{CodeSpecial}, %
	keywordstyle=[3]\color{CodeSpecial},   %
	keywordstyle=[4]\bfseries\color{CodeKeyword},  %
	keywordstyle=[5]\color{CodeSpecial}, %
	keywordstyle=[6]\bfseries, %
	commentstyle=\color{CodeComment},
	stringstyle=\color{CodeString},
	columns=spaceflexible,
	keepspaces=true,
	showspaces=false,
	showtabs=false,
	showstringspaces=false,
}

\newtcblisting{viperblock}{
	enhanced,
	colback=CodeBack,
	colframe=CodeBack, %
	sharp corners,
	top=0pt, bottom=0pt,
	left=2pt, right=2pt,
	listing only,
	listing options={
		language=Viper,
		style=colouredViper,
		frame=none,
		numbers=none,
		backgroundcolor=,
		rulecolor=,
		framerule=0pt,
		xleftmargin=0pt,
		xrightmargin=0pt,
		aboveskip=0pt,
		belowskip=0pt,
		basicstyle=\ttfamily\small,
		mathescape=true,
		escapechar=~,
		tabsize=2,
	}%
}{}

\lstdefinelanguage{Rust}
{
	sensitive,
	morekeywords=[1]{
		&,*,
		extern,
		unsafe,async,await,move,
		use,pub,crate,super,self,Self,mod,
		struct,enum,fn,const,static,let,mut,ref,type,impl,dyn,trait,where,as,
		break,continue,if,else,while,for,loop,match,return,yield,in,
		loc,box,strg,
		&loc,&mut,&strg,
		pack,unpack,unpacking,forget,remember,transfer,def,in,
	},
	morekeywords=[2]{
		assert!, assert_eq!, debug_assert!, debug_assert_eq!, debug_assert_ne!,
		panic!, unimplemented!, unreachable!,
		print!, println!, vec!, matches!
	},
	morekeywords=[3]{
		pure, trusted, ensures, requires, result, invariant, extern_spec,
		ghost_fn, pure_memory, pure_unstable,
        logic, specs, %
		body_invariant!, predicate!,
		forall, old, exists,
		old, after_expiry, before_expiry, at_expiry, on_expiry, assert_on_expiry,
		owns, capable,
		readRef, writeRef, localRef, local, unique, write, immutable, read, noReadRef, noWriteRef,
		readRefCall, writeRefCall, localRefCall, localCall, uniqueCall, writeCall, noReadRefCall, noWriteRefCall,
		readRefStmt, writeRefStmt, localRefStmt, localStmt, uniqueStmt, writeStmt, noReadRefStmt, noWriteRefStmt,
	},
	morekeywords=[4]{
		bool,u8,u16,u32,u64,u128,i8,i16,i32,i64,i128,char,str,usize
	},
	morekeywords=[5]{
	},
	morekeywords=[6]{
		true, false
	},
	morecomment=[l]{//},
	morecomment=[s]{/*}{*/},
	morecomment=[l]{///},
	morecomment=[s]{/*!}{*/},
	morecomment=[l]{//!},
	string=[b]{"},
	string=[s]{f"}{"},
	alsoletter={!},
}

\lstdefinestyle{colouredRust}
{
	basicstyle=,
	identifierstyle=,
	keywordstyle=[1]\bfseries\color{CodeKeyword},  %
	keywordstyle=[2]\color{CodeSpecial}, %
	keywordstyle=[3]\color{CodeSpecial},   %
	keywordstyle=[4]\bfseries\color{CodeKeyword},  %
	keywordstyle=[5]\bfseries\color{CodeKeyword}, %
	keywordstyle=[6]\bfseries, %
	commentstyle=\color{CodeComment},
	stringstyle=\color{CodeString},
	literate={!matches!}{!{\color{CodeSpecial}matches!}}1,
	columns=spaceflexible,
	keepspaces=true,
	showspaces=false,
	showtabs=false,
	showstringspaces=false,
}

\newcommand{\rust}[1]{%
	\text{\lstinline[
		language=Rust,
		style=colouredRust,
		backgroundcolor=\color{CodeBack},
		basicstyle=\ttfamily, %
		breaklines,
		breakatwhitespace,
		mathescape=true,
		escapechar=~,
	]|#1|}%
}

\newtcblisting{rustblock}{
	colback=CodeBack,
	colframe=CodeBack, %
	sharp corners,
	top=0pt, bottom=0pt,
	left=2pt, right=2pt,
	listing only,
	listing options={
		language=Rust,
		style=colouredRust,
		frame=none,
		numbers=none,
		backgroundcolor=,
		rulecolor=,
		framerule=0pt,
		xleftmargin=0pt,
		xrightmargin=0pt,
		aboveskip=0pt,
		belowskip=0pt,
		basicstyle=\ttfamily\small,
		mathescape=true,
		escapechar=~,
		tabsize=2,
	},
}

\newcommand{\ie}{i.e.}
\newcommand{\eg}{e.g.}

\newcommand{\etc}{etc.}

\newcommand*{\labtab}[1]{\label{tab:#1}}
\newcommand*{\labsec}[1]{\label{sec:#1}}
\newcommand*{\labsubsec}[1]{\label{subsec:#1}}
\newcommand*{\labsubsubsec}[1]{\label{subsubsec:#1}}

\newcommand*{\tabref}[1]{Table~\ref{tab:#1}}
\newcommand*{\figref}[1]{Fig.~\ref{fig-#1}}

\newcommand*{\secref}[1]{Sec.~\ref{sec:#1}}
\newcommand*{\subsecref}[1]{Sec.~\ref{subsec:#1}}

\newcommand*{\sectionref}[1]{Section~\ref{sec:#1}}

\newcommand*\circled[1]{%
    \tikz[baseline=(char.base)]{%
        \node[shape=circle,draw,inner sep=1pt] (char) {\smaller #1};%
    }%
}

\newcounter{keyproblemcounter}

\newcounter{challengecounter}

\newcommand*{\ang}[1]{\ensuremath{\langle{}\text{#1}\rangle{}}}

\pgfdeclarelayer{nodelayer}
\pgfdeclarelayer{edgelayer}
\pgfsetlayers{edgelayer,main,nodelayer}

\tikzstyle{kind}=[fill=white, draw=black, shape=rectangle, rounded corners=5pt, line width=0.75pt]

\tikzstyle{implication}=[-Latex, line width=1pt]
\tikzstyle{incompatibility}=[draw=red, dashed, -, line width=1.5pt]

\newenvironment{closealign}{%
	\par\centering$\displaystyle\begin{aligned}%
}{%
	\end{aligned}$\par%
}

\newcommand{\writeRef}{\textbf{writeRef}}
\newcommand{\readRef}{\textbf{readRef}}
\newcommand{\readC}{\textbf{read}}
\newcommand{\writeC}{\textbf{write}}
\newcommand{\immutable}{\textbf{immutable}}
\newcommand{\unique}{\textbf{unique}}
\newcommand{\local}{\textbf{local}}
\newcommand{\noReadRef}{\textbf{noReadRef}}
\newcommand{\noWriteRef}{\textbf{noWriteRef}}

\begin{document}

\title{Reasoning about Interior Mutability in Rust using  Library-Defined Capabilities}

\author[F. Poli]{Federico Poli}
\orcid{0000-0001-7709-1965}
\affiliation{
  \department{Department of Computer Science}
  \institution{ETH Zurich}
  \country{Switzerland}
}
\email{federico.poli@inf.ethz.ch}

\author[X. Denis]{Xavier Denis}
\orcid{0000-0003-2530-8418}
\affiliation{
  \department{Department of Computer Science}
  \institution{ETH Zurich}
  \country{Switzerland}
}
\email{denis.xavier@inf.ethz.ch}

\author[P. M\"uller]{Peter M\"uller}
\orcid{0000-0001-7001-2566}
\affiliation{
  \department{Department of Computer Science}
  \institution{ETH Zurich}
  \country{Switzerland}
}
\email{peter.mueller@inf.ethz.ch}

\author[A. J. Summers]{Alexander J. Summers}
\orcid{0000-0001-5554-9381}
\affiliation{
  \department{Department of Computer Science}
  \institution{University of British Columbia}
  \country{Canada}
}
\email{alex.summers@ubc.ca}

\date{2024}

\begin{abstract}
Existing automated verification techniques for safe Rust code rely on the strong type-system properties to reason about programs, especially to deduce which memory locations do not change (\ie, are framed) across function calls.
However, these type guarantees do not hold in the presence of \emph{interior mutability} (\eg, when interacting with any concurrent data structure). 
As a consequence, existing verification techniques for safe code such as Prusti~\cite{prusti} and Creusot~\cite{creusot} are either unsound or fundamentally incomplete if applied to this setting.
In this work, we present the first technique capable of automatically verifying safe clients of existing interiorly mutable types.
At the core of our approach, we identify a novel notion of \emph{implicit capabilities}: library-defined properties that cannot be expressed using Rust's types. 
We propose new annotations to specify these capabilities and a first-order logic encoding suitable for program verification.
We have implemented our technique in a verifier called Mendel and used it to prove absence of panics in  Rust programs that make use of popular standard-library types with interior mutability, including \rust{Rc}, \rust{Arc}, \rust{Cell}, \rust{RefCell}, \rust{AtomicI32}, \rust{Mutex} and \rust{RwLock}.
Our evaluation shows that these library annotations are \emph{useful} for verifying usages of real-world libraries, and \emph{powerful} enough to require zero client-side annotations in many of the verified programs.
\end{abstract}


\maketitle

\section{Introduction}\labsec{04-introduction}

Rust's ownership type system offers strong guarantees, such as memory safety, absence of data races, absence of dangling pointers, and, in general, absence of undefined behavior (UB). 
In the safe language fragment of Rust, these properties are statically guaranteed by the compiler, making it possible for verification techniques and tools to build upon them~\cite{prusti, creusot, verus, aeneas}.
This is achieved by associating an \emph{exclusive capability}~\cite{capabilities-for-sharing} to all mutable references and non-borrowed types and a \emph{shared capability} to all immutable references.
The former capability guarantees write access and non-aliasing, while the latter read access and immutability. 

The \emph{explicit} capabilities prescribed by Rust types are too restrictive to implement certain behaviors: cyclic data structures such as doubly linked lists, concurrent data structures, shared mutable state such as a global cache, etc.
To overcome this limitation, Rust libraries may use unsafe Rust in their implementation to mutate state reached via shared references. 
Such libraries exhibit \emph{interior mutability} in their API\@.

However, the additional expressivity for such library developers comes at the cost of losing static guarantees. 
In the presence of interior mutability, the explicit capabilities of an API no longer describe all mutations potentially performed by a library and, thus, no longer provide the static information needed to verify clients.
This is an inherent limitation of existing modular verification tools, none of which can reason about basic usages of standard library types with interior mutability.

The program in \figref{04-introduction-example-cell} shows a simple usage of the \rust{Cell} type of Rust's standard library, a container that uses interior mutability to provide an aliasable mutable cell in Rust, and thus can be mutated via calls made on shared references. 
The \rust{cell_client} function receives a shared reference \rust{c} pointing to an instance of \rust{Cell}. 
We use \rust{c.set(..)} to increment the cell's content.
The two \rust{c.get()} calls around the increment return copies of the cell's content.
At the end, with an \rust{assert!(..)} statement, the function checks at runtime whether the value read after the increment is precisely one more than the value read before.

The assertion never fails at runtime because the design of the \rust{Cell} library guarantees that its content cannot be modified concurrently by another thread as long as there is a \rust{&Cell} instance like \rust{c}, so the only modification happens as a result of the \rust{set} call. 
However, existing verification techniques for Rust do not use this information.
To soundly model the interior mutability of \rust{Cell}, they model it as an unknown value that changes unpredictably and conservatively, assume that other threads might interfere at any moment and modify the cell's content. 
Consequently, existing verifiers cannot prove the last assertion, for which they report a spurious verification error.

\begin{figure}[htb]
\begin{rustblock}
fn cell_client(c: &Cell<i32>) {
  let before = c.get(); c.set(before + 1); let after = c.get();
  assert!(before + 1 == after); // Goal: prove that this never fails
}
\end{rustblock}
\caption{Example of a client of the \rust{Cell} library.}
\label{fig-04-introduction-example-cell}
\end{figure}

This work presents the first verification technique that enables automated and precise reasoning about clients of interiorly mutable types. 
At the core of our work, we identify a novel notion of \emph{implicit capabilities}, which are properties of library types that express guarantees provided by the library design rather than Rust's types. 
By leveraging these capabilities, our technique can effectively frame and reason about interior mutability.
We implemented our technique in a verifier called \textbf{Mendel}\footnote{Gregor Mendel, the father of modern genetics, opened the way to the study of \emph{capabilities} and \emph{mutations} of DNA.}, the first tool capable of automatically verifying code such as \figref{04-introduction-example-cell}.

Note that our technique focuses on modularly verifying \emph{clients} of interiorly mutable libraries using trusted library annotations. 
Verifying the libraries is orthogonal and requires different techniques to handle their unsafe code.

\paragraph{Contributions}
The main contributions of our work are:
\begin{enumerate}
	\item We identify the notion of implicit capabilities, define a rich family of these and propose new annotations to specify them on library types.
	\item We present a new verification technique that uses capabilities to reason about safe Rust code, even in the presence of interior mutability.
	\item We define an encoding of our reasoning technique to first-order logic, suitable for automation using an SMT-based verification toolchain.
	\item We implement our reasoning technique in a deductive verifier for Rust called Mendel, showing with an evaluation that our technique is (1)~\emph{useful}, as it supports popular types with interior mutability defined in the standard library and (2)~\emph{lightweight} because it requires near-zero annotations on the client side of the simple programs in our evaluation.
\end{enumerate}

\paragraph{Outline}
The rest of this paper is structured as follows.
\sectionref{04-problems} presents the verification challenges of our setting and identifies our novel notion of implicit capability.
\sectionref{sec-overview} gives an overview of our technique and tool from a user's perspective.
In \sectionref{04-approach}, we present our implicit capabilities, introducing library annotations to specify them and semantic rules to reason about them. 
We illustrate our technique using examples in \sectionref{04-approach-examples}.
\sectionref{04-soundness} introduces our novel validation technique to derive properties of sound Rust abstractions from well-established properties of safe Rust.
\sectionref{04-encoding} presents a first-order logic encoding of our capability-based reasoning technique.
\sectionref{04-evaluation} discusses the implementation of our technique in a deductive verification tool and its evaluation on clients of popular standard-library types.
\sectionref{04-related-work} discusses related techniques, and we conclude in \sectionref{04-conclusion}.

\section{Verification Challenges}\labsec{04-problems}

In this section, we present the key problems of verifying clients of libraries exposing interior mutability.

\subsection{Shared Mutable State}

Interior mutability provides an essential escape hatch from Rust's ownership discipline. 
It plays a central role in synchronization primitives such as \rust{Mutex} and for memoization in types like \rust{RefCell}. 
At the same time, the aliased mutable state often implied by interior mutability complicates verification. 
Neither verification techniques that view (safe) Rust programs as essentially functional code~\cite{creusot, verus, aeneas} nor techniques based on separation logic~\cite{prusti} have a built-in support for shared mutable state. 
In principle, these techniques could be extended to perform some form of rely-guarantee reasoning~\cite{deny-guarantee,Dinsdale-YoungDGPV10}, which would require defining an invariant and protocol for each mutably aliased region and verifying that all reads and writes to that region maintain the protocol.

Rely-guarantee reasoning is powerful, but overly complicated for the usage patterns of shared mutable state in many commercially used Rust libraries. 
In contrast to traditional imperative languages, Rust's library types typically constrain their usage of interior mutability; they provides a safe interface with specific guarantees about the aliasing and mutability.
These \emph{implicit capabilities} cannot be checked by the Rust compiler, but they provide powerful informal reasoning principles to Rust programmers. 
For example, given an object \rust{c : &Cell}, calling \rust{c.get()} twice in a row is guaranteed to return the same result, a consequence of the single-threaded nature of \rust{Cell}. 
Existing verifiers cannot leverage these capabilities and thus struggle to specify and verify this property. 
This brings us to our first challenge: \textbf{how could a verifier leverage the implicit capabilities of Rust types to reason about interior mutability?}

\subsection{Usability}

All existing verification techniques for safe Rust enable programmers to reason at the abstraction level of the Rust program without exposing them to complex program logic. 
Our goal is to retain this abstraction also in the presence of interior mutability. 
To achieve this goal, the guarantees provided by Rust libraries should be expressed via simple annotations rather than complex logical formulas. 
The challenge, thus, is \textbf{how to specify implicit capabilities in a way that is simple for programmers but provides enough information to enable precise reasoning?}

\subsection{Conditional Library Properties}

Not every usage of interior mutability is as simple as \rust{Cell} or \rust{Mutex}: the \rust{Arc} type of atomic reference counted pointers uses interior mutability to update its reference count. 
When exactly one copy of the \rust{Arc} is around, it becomes possible to safely assume the value is unchanged between two consecutive reads; no other thread could have modified this unique \rust{Arc}. 
This capability is \emph{conditional} on the runtime value of the reference counter, demonstrating a need for first-class support for capabilities in specifications: \textbf{we should be able to condition capabilities on arbitrary runtime expressions}.

\section{A whirlwind tour of Mendel}
\labsec{sec-overview}

We implemented our technique in a deductive verifier called 
Mendel. 
Reasoning about interior mutability is based on a novel lightweight \emph{capability annotation}.
These capabilities describe guarantees of individual types that are not captured by Rust's type system. 
Mendel can leverage capabilities to automatically reason about the correctness of subtle Rust programs, as we illustrate in this section.

\begin{figure*}
\begin{minipage}{0.59\linewidth}
\begin{rustblock}
#[extern_spec]
#[capable(&self => local(self.as_ptr()))] ~\circled{A}~
impl<T : Copy> Cell<T> {
  #[pure_memory]
  pub fn as_ptr(&self) -> *mut T;

  #[pure_unstable] ~\label{cell:puremem}~
  #[ensures(result ==== deref(self.as_ptr()))] ~\circled{B$_1$}~
  pub fn get(&self) -> T;

  #[ensures(deref(self.as_ptr()) ==== value)] ~\circled{B$_2$}~
  pub fn set(&self, value: T);
}
\end{rustblock}
\end{minipage}
\hfill
\begin{minipage}{0.39\linewidth}
\begin{rustblock}
fn cell_client(c: &Cell<i32>) {
  let before = c.get();
  c.set(before + 1);
  let after = c.get();
  // Ok!
  assert!(before + 1 == after);
}
\end{rustblock}
\end{minipage}
\label{fig-overview-cell}
\caption{The motivating example from \Cref{fig-04-introduction-example-cell}, together with an annotated version of the \rust{Cell} API\@. The annotations express implicit capabilities of the type, as well as postconditions and purity for its functions.}
\end{figure*}

\subsection{Motivating Example}

We start with the motivating examples from \Cref{fig-04-introduction-example-cell}. 
We restated the code and the required Mendel specification in \Cref{fig-overview-cell}.
On the left side we show the (trusted) annotations of the \rust{Cell} API, provided as part of the Mendel tool. 
Line~\circled{A} attaches a \emph{local} capability to \rust{&Cell<T>}. 
The \local{} capability ensures that as long as a \rust{&Cell<T>} reference exists, all reads must occur from the same thread. 
We also provide specifications for \rust{get} and \rust{set} on lines \circled{B$_1$} and \circled{B$_2$} which describe their functional behavior. 
The annotations \rust{pure_memory} and \rust{pure_unstable} denote different levels of `purity' which are available for functions in Mendel. 
All pure functions are side-effect free and deterministic; he different purity annotations express which parts of the heap they may depend on, as we explain later.

Given these specifications, Mendel verifies \rust{cell_client} automatically.
Because \rust{c} is of type \rust{&Cell<i32>}, Mendel can use the \local{} capability to conclude that only the code of \rust{cell_client} can modify the cell's value while the function executes and, thus, that the value of \rust{after} is exactly the one provided to \rust{set}. 
By combining this local reasoning with the functional specifications for \rust{get} and \rust{set}, Mendel proves that the assertion holds.

As this example shows, verification for clients of \rust{Cell} is lightweight and requires no annotations on the client code. 
As usual in deductive verifiers, the annotations on the library are trusted and provided by the tool developers.

\subsection{Lightweight Non-aliasing}

\begin{figure*}
  \begin{rustblock}
  #[capable(&mut self => writeRef(self.as_ptr()))] 
  impl<T> RefCell<T> {}
  
  #[capable(&self => readRef(self.refcell().as_ptr()))]
  impl<'b, T> Ref<'b, T> {}
  
  fn use_refcell(x: &RefCell<i32>) { /* omitted */ }
  
  fn refcell_client(x: &RefCell<i32>, y: RefMut<i32>) {
    let Ok(a /*: Ref */) = x.try_borrow() else { return; };
    let before: i32 = *a;
    use_refcell(x);
    let Ok(b /*: Ref */) = x.try_borrow() else { return; };
    let after: i32 = *b;
    assert!(before == after);  // Both assertions succeed
    assert!(x.as_ptr() as *const _ != y.deref() as *const _);
  }
  \end{rustblock}
  \caption{Example of a client of the \rust{RefCell} library. The full annotations on the \rust{RefCell} library can be seen in the test suite of the evaluation.}
  \label{fig-04-approach-core-refcell}
  \end{figure*}

Rust's \rust{RefCell} type provides mutable memory locations with dynamically checked borrow rules. 
The capability annotations in \figref{04-approach-core-refcell} state two essential guarantees of the \rust{RefCell} and \rust{Ref} libraries: (1)~Any \rust{&mut RefCell} instance is capable of obtaining a mutable reference to its content.
That is, mutable instances of \rust{RefCell} hold the unique capability for mutating their content. 
This annotation uses the existing \rust{RefCell::as_ptr} API method to identify the content of the \rust{RefCell} by address.
(2)~Any \rust{&Ref} instance is capable of obtaining a shared reference to the content of the associated \rust{RefCell}. 
To formalize this property, the annotation uses an auxiliary specification method \rust{Ref::refcell} (not shown in the figure) that models the \rust{RefCell} instance associated with a \rust{Ref}.
  
Using these two annotations, Mendel verifies panic-freedom of the client function \rust{refcell_client} as follows.
First, we observe that \rust{a} is alive across the \rust{use_refcell(x)} call.
From the \readRef{} capability of \rust{a}, we can obtain an \immutable{} capability, which allows us to establish that \rust{a} is unchanged across calls. 
This, combined with standard functional specifications for the other methods in the program, makes it possible to verify the first assertion.
  
Second, before the last assertion, the instances \rust{a} and \rust{y} are both alive. 
The instance \rust{a} still holds an \immutable{} capability that implies immutability of \rust{x}'s content. 
According to the first capability annotation in \figref{04-approach-core-refcell}, the instance \rust{y} of type \rust{RefCell} holds an implicit capability that implies exclusive mutable access to \rust{y}'s content. 
Since these two capabilities are incompatible, it follows that the content of \rust{x} \emph{must} be at a different location than the content of \rust{y}, which proves the second and last assertion. 
Without the library capability annotations, none of the two assertions could be verified.

\subsection{Conditional Capabilities}

Types like the atomic reference counter \rust{Arc} use interior mutability to modify their counters. 
The current count affects the result of different API functions like \rust{Arc::into_inner}\footnote{\url{https://doc.rust-lang.org/std/sync/struct.Arc.html\#method.into_inner}} which succeeds only if there is a single outstanding reference.
The example in \Cref{fig-04-problems-example-arc} tests the count of \rust{x} to determine whether it is safe to call \rust{into_inner}. 
If the strong count is exactly one, then we know that this count is stable, that \rust{x} and \rust{y} do not alias (since \rust{x} must be the only copy of this reference-counted pointer), and that the call to \rust{into_inner} will succeed (for the same reason).
However, if our test determined that the count was not 1, none of these assertions hold: the reference count could have been decremented by another thread to 1 after the test, and \rust{x} could alias \rust{y}.

Mendel can perform these reasoning steps based on the annotation on line \circled{A} which grants a \local{} capability \emph{only if} the count is exactly 1. 
This capability allows Mendel to verify the then-branch in \rust{client} by ruling out  interference from other threads (but not the else-branch, where the condition of the capability does not hold.

\begin{figure*}
  \begin{minipage}{0.59\linewidth}

  \begin{rustblock}
  fn client(mut x: Arc<i32>, y: Arc<i32>) -> i32 {
    if x.strong_count() == 1 { // All 3 succeed
      assert!(x.strong_count() == 1); 
      assert!(x.as_ptr() != y.as_ptr()); 
      Arc::into_inner(x).unwrap()
    } else { // All 3 assertions fail
      assert!(x.strong_count() != 1);
      assert!(x.as_ptr() != y.as_ptr());
      assert!(x.into_inner().is_none());
      0
    }
  }
  \end{rustblock}
\end{minipage}
\hfill
\begin{minipage}{0.39\linewidth}
\begin{rustblock}
#[extern_spec]
#[capable(&self 
  if self.strong_count() == 1,~\circled{A}~
  local(self.as_ptr())
)]
impl<T> Arc<T> {
  // Functional specs for Arc<T>
}
\end{rustblock}
\end{minipage}
  \caption{Example of a client of the \rust{Arc} library. The then-branch is correct, while all assertions in the else-branch might fail. 
  Mendel verifies the then-branch successfully, using the conditional capability of the \rust{Arc} type.}
  \label{fig-04-problems-example-arc}
  \end{figure*}

\section{Implicit Capabilities and Framing}\labsec{04-approach}

In this section, we present the syntax of our capabilities, illustrate their usage on the example from \figref{04-approach-core-refcell}, and
present derived properties of our capabilities that enable additional reasoning steps.

\newcommand{\axiom}[1]{\inferrule{}{#1}}

\subsection{Syntax of Capabilities}
\labsubsec{approach-syntax}

Our technique supports the following capabilities (the implementation provides additional syntactic constructs, which are not essential here).

\begin{adjustbox}{valign=t,minipage={0.49\textwidth}}
\begin{align*}
\text{cap} \ni c &\bnfdef{} \readRef(p) \mid{} \writeRef(p) 
	\mid{} \readC(p) \\ &\mid{} \writeC(p) \mid{} 
	\immutable(p) \\&\mid{} \unique(p) \mid{} \local(p) \\
	&\mid{} \noReadRef(p) \mid{} \noWriteRef(p) \mid{} c_1 \land c_2 \\
\end{align*}
\end{adjustbox}
\hfill
\begin{adjustbox}{valign=t,minipage={0.49\textwidth}}
\begin{align*}
\text{place} \ni p &\bnfdef{} x \mid \textbf{$\ast$}~p \mid p\textbf{.}\textit{f} \\
\text{type} \ni \tau &\bnfdef{} \textbf{int} \mid{} \textbf{bool} \mid{} \textbf{\&mut}~\tau \\
	& \mid \textbf{\&}~\tau \mid{} \tau_1 \times \tau_2 \mid{} \tau_1 + \tau_2\\
\text{contract} &\bnfdef{} \textbf{capable}(\tau~\textbf{if}~e \Rightarrow c) \\
\text{expr} \ni e &\bnfdef{} \textbf{true} \mid{} \textbf{false} \mid{} x \mid{} e_1 \land e_2 \mid{} ... \\
\end{align*}
\end{adjustbox}

Each capability $c$ falls into one of the following categories:
\begin{enumerate}
	\item \textbf{Core capabilities}, corresponding to the explicit capabilities of Rust references. 
	\readRef{} corresponds to the capability of shared references. 
	It provides \emph{shared} read-only access, with the guarantee that the target memory location is not modified via aliases.
	\writeRef{} corresponds to the capability of mutable references and of fully initialized non-borrowed types. 
	It provides \emph{exclusive} read and write access.

	\item \textbf{Fine-grained capabilities} to read, write, express immutability, unique access, or thread-local ownership (\readC, \writeC, \immutable, \unique, \local).
	These are fundamental properties used across many program reasoning techniques, such as the type system of Pony~\cite{pony}, experimental type systems for C\#~\cite{uniqueness-immutability} and, in general, separation logic~\cite{separation-logic}.
	Compared to the core capabilities, \readC and \writeC are weaker definitions that lack the immutability and non-aliasing guarantees of shared and mutable references.
	Likewise, \unique{} is weaker than \writeRef, but for a subtle reason that we will discuss in \secref{04-soundness}: \writeRef{} implies that it is possible to obtain a mutable reference through the library API, while \unique{} does not.
	As the name implies, \immutable{} ensures the target does not change its value across any statement.

	\item \textbf{Deny capabilities} \noReadRef{} and \noWriteRef
	 express that there \emph{cannot} exist references to a certain memory location. 
	This makes it possible to deduce (1)~non-aliasing properties between existing references used in safe Rust and memory locations managed by libraries and (2)~immutability of memory locations across assignments to mutable references, as done to verify the example in \figref{04-approach-core-refcell}.
\end{enumerate}

Each capability is indexed by a place $p$, which is a syntactic representation of a memory location.
Capabilities are attached to types through $\textbf{capable}(..)$ annotations, which can optionally include a condition $e$ that must hold for the capability to be valid.
When the condition is $true$, we write $\textbf{capable}(c)$ instead.

\subsection{Reasoning with Capabilities}
\labsubsec{approach-derived}

During verification, our technique manipulates capabilities in three major ways: (1)~It derives from the capabilities of a composite type the capabilities of its components, for instance, the capabilities of a field $p.f$ from the capabilities of a reference $p$. 
(2)~It derives capabilities from others for the same place, for instance, a $\readRef$ capability from a $\writeRef$ capability. 
(3)~It exploits incompatibilities between capabilities; for instance, $\writeRef(p)$ and $\writeRef(q)$ imply that $p$ and $q$ are not aliased. 
In the following, we formalize these important properties of capabilities.

\begin{figure}
	\begin{mathpar}
	\writeRef(p) \vdash \writeRef(p.f)

	\readRef(p) \vdash \readRef(p.f)

	\inferrule{p : \textbf{\&mut}~T}
	{ \writeRef(p) \vdash \writeRef(*p)}

	\inferrule{p : \textbf{\&}~T \; \lor \; p : \textbf{\&mut}~T}
	{ \readRef(p) \vdash \readRef(*p)}
	\end{mathpar}
	\caption{Structural properties of capabilities.}
	\label{fig-04-approach-core-structural}
\end{figure}

The structural rules in \figref{04-approach-core-structural} allow us transform a capability of a reference type into a capability of the pointee type, a capability of a product type into capabilities of its components, and so on. We present here the rules for $\readRef(p)$ and $\writeRef(p)$, but analogous rules exist for other capabilities.

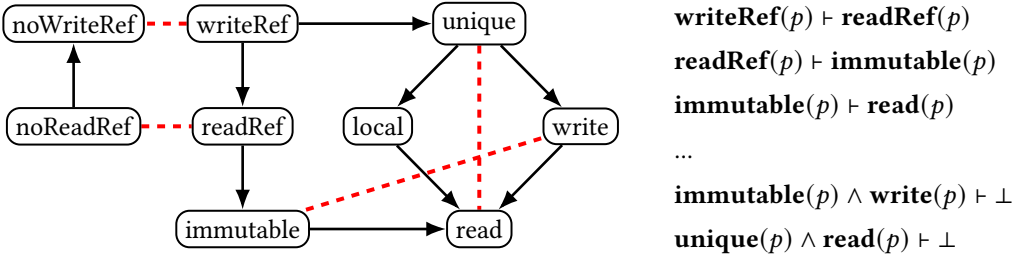
\begin{figure}
	\begin{adjustbox}{valign=t,minipage={0.59\textwidth}}
		\begin{tikzpicture}[scale=0.9]
	\begin{pgfonlayer}{nodelayer}
		\node [style=kind] (1) at (-5.5, 1.5) {noWriteRef};
		\node [style=kind] (0) at (-5.5, 0) {noReadRef};
		\node [style=kind] (2) at (-3, 1.5) {writeRef};
		\node [style=kind] (5) at (-3, 0) {readRef};
		\node [style=kind] (8) at (-3, -1.5) {immutable};
		\node [style=kind] (4) at (0.5, 1.5) {unique};
		\node [style=kind] (6) at (-1, 0) {local};
		\node [style=kind] (7) at (2, 0) {write};
		\node [style=kind] (9) at (0.5, -1.5) {read};
	\end{pgfonlayer}
	\begin{pgfonlayer}{edgelayer}
		\draw [style=implication] (2) to (5);
		\draw [style=implication, anchor=south west] (2) to (4); %
		\draw [style=implication] (0) to (1);
		\draw [style=implication] (4) to (6);
		\draw [style=implication] (4) to (7);
		\draw [style=incompatibility] (0) to (5);
		\draw [style=incompatibility, in=0, out=-180] (2) to (1);
		\draw [style=implication] (5) to (8);
		\draw [style=implication] (8) to (9);
		\draw [style=implication] (6) to (9);
		\draw [style=implication] (7) to (9);
		\draw [style=incompatibility] (4) to (9);
		\draw [style=incompatibility] (8) to (7);
	\end{pgfonlayer}
\end{tikzpicture}
	\end{adjustbox}
	\hfill
	\begin{adjustbox}{valign=t,minipage={0.39\textwidth}}
		\begin{closealign}
			&\axiom{\writeRef(p) \vdash \readRef(p)} \\
			&\axiom{\readRef(p) \vdash \immutable(p)} \\
			&\axiom{\immutable(p) \vdash \readC(p)} \\
			&\axiom{...} \\
			&\axiom{\immutable(p) \land \writeC(p) \vdash \bot} \\
			&\axiom{\unique(p) \land \readC(p) \vdash \bot}
		\end{closealign}
	\end{adjustbox}
	\caption{Diagram (on the left) of the extended capability implications (arrow edges) and incompatibilities (dashed edges), with their mathematical interpretation on the right (implications, then incompatibilities).
	}
	\label{fig-04-approach-ext-capability-kinds}
\end{figure}

The implications and incompatibilities between capabilities for the same place are shown in \figref{04-approach-ext-capability-kinds} by arrows and dashed edges, respectively. 
Implications are transitive, but incompatibilities are not. 
Note that $\readC(p)$ and $\writeC(p)$ are not incompatible; in fact, types like \rust{Cell} rely on having both capabilities for the same memory location.

\paragraph{Framing Properties}\labsubsubsec{04-approach-ext-immutability}

As we illustrated with the examples in the previous section, verification uses capabilities to derive non-aliasing properties and framing properties. 
The former follows from the incompatibilities discussed above.
To derive the latter, we apply the capability reasoning \emph{across} each statement as we explain next.

We say a capability is held \emph{across} a statement if it is available both before and after a statement. 

We can now define the framing properties that Mendel uses to automate reasoning:
\begin{itemize}
	\item An $\immutable(p)$ capability guarantees that $p$ can be framed out of any statement $S$. 
	\item 
	If $p$ does not occur in a statement $S$, then a $\unique(p)$ capability allows excluding $p$ from the frame. 
	\item For any assignment $q = e$, the combination of $\local(p)$ and $\noWriteRef(p)$ ensure that $p$ can be framed around the assignment.  
	The $\local$ capability ensures other threads cannot modify $p$ while the $\noWriteRef$ ensures the current thread cannot either. 
	\item 
	Across call statements $q = f(\bar{x})$ of functions that are marked by a purity annotation (presented in \subsecref{04-approach-purity}), a $\local(p)$ and $\noWriteRef{p}$ capability allows framing $p$ across the call.
\end{itemize}

\subsection{Purity Annotations and Framing of Function Values}\labsubsec{04-approach-purity}

Libraries typically offer methods to inspect data structures, for instance, to obtain the length of a vector. 
When such functions are deterministic, side-effect free, and always terminate, we call them \emph{pure} and allow them to be used in specifications. 
The result of a pure method typically depends on the state of their arguments.
For instance, \rust{Vec::len} returns the length of a vector by reading its value from a private field.

Reasoning about the results of pure methods, particularly the framing properties of their results in the presence of heap modifications, requires information about the memory locations on which a pure method's result depends. 
In the absence of unsafe code, framing is simple: calling a pure method on equal arguments will lead to equal results.
However, this conclusion might be invalid when pure methods depend on memory locations subject to interior mutability because these locations could be modified through an alias.

Among libraries with interior mutability, \rust{Cell::as_ptr} is a pure method that returns a raw pointer whose target address is computed as a fixed offset from the address where the \rust{Cell} instance is stored.
Consequently, framing is not compromised by interior mutability.
In the same API, the \rust{Cell::get} method is pure as well because it returns the value contained in the \rust{Cell} instance by internally dereferencing the result of \rust{Cell::as_ptr}. 
As discussed earlier, mutations via aliases might affect the result of \rust{Cell::get}, but only within the same thread.
However, the result of \rust{Vec::len} is only affected by the vector's contents; the same vector can be moved in memory without affecting its result.

These three methods are all pure but provide different framing guarantees. 
In the interest of lightweight specifications, we abstain from equipping pure methods with precise specifications of their read effects. 
Instead, we define three classes of pure functions --- \rust{#[pure]}, \rust{#[pure_memory]}, and \rust{#[pure_unstable]} --- corresponding to different footprint definitions, ordered from the more to the less restrictive.

Methods declared \rust{#[pure]} (such as \rust{Vec::len}) may depend only on the \emph{values} reachable from its arguments. 
These values might be reached by following fields, dereferencing references, or calling other \rust{#[pure]} functions. 
Crucially, pure-value methods must \emph{not} depend on the content of types with interior mutability (\ie, \rust{UnsafeCell}). 
Moreover, pure-value methods may depend on the address of the target of a raw pointer but not on the value of its target. 
Conversely, they may depend on the value of the target of a reference but not on the address of their target.
This follows the intuition that the value of a raw pointer is its target address, while the value of a reference is the value of its target. 
These rules provide strong framing properties; interior mutability does not affect pure-value methods.

Methods declared \rust{#[pure_memory]} (such as \rust{Cell::as_ptr}) may depend on the values reachable from its arguments (like pure-value methods) and, additionally, on their memory addresses. 

The difference between \rust{#[pure]} and \rust{#[pure_memory]} functions is that the result of the former is guaranteed to be the same no matter how values are moved in memory, while the latter can observe (and return) the different memory addresses reachable from its parameters.

Methods declared \rust{#[pure_unstable]} (such as \rust{Cell::get}) may depend on any memory value, including the content of types with interior mutability. 
Consequently, concurrent modifications might affect the result of such methods. 
This purity level does not enable any form of framing; programmers must provide postconditions that effectively restrict the footprint.
For example, as shown in \figref{04-approach-ext-cell-lib} below, the postcondition of \rust{Cell::get} states that the result is equal to the dereference of \rust{self.as_ptr()}, where \rust{====} expresses structural equality that considers both values and addresses (\eg, the target address of references).
In contrast, \rust{deref} is a built-in pure-unstable function that models dereferences. 
The capabilities that \rust{Cell} declares for the memory location \rust{self.as_ptr()} make it possible to reason about the results of multiple \rust{Cell::get} calls.
In pure contexts, the results of pure-unstable functions with the same arguments are guaranteed to be evaluated to the same value.

The purity of method implementations is checked syntactically. 
These checks ensure, among other rules, that pure methods have copy-type parameters, call only pure methods of the appropriate purity level, contain assignments only to local variables, and follow the rules of the different purity levels explained above.

\section{Case Studies}\labsec{04-approach-examples}

By using the capability and purity annotations presented so far, it is now possible to explain the main proof steps needed to verify the examples in \figref{04-problems-example-refcell}, and \figref{04-problems-example-arc}, which make use several types of the standard library. 
The full details of the encoding and memory model will be presented in \secref{04-encoding}, while the full annotation of the library types is available in the artifact of the evaluation (\secref{04-evaluation}).

\subsubsection{Usage of Cell}

\begin{figure*}[htb]
\begin{rustblock}
#[extern_spec]
#[capable(&self => local(self.as_ptr()))]
#[capable(&self => noReadRef(self.as_ptr()))]
#[capable(&self => noWriteRef(self.as_ptr()))]
#[capable(&mut self => writeRef(self.as_ptr()))]
impl<T> Cell<T> {
  #[pure_memory]
  pub fn as_ptr(&self) -> *mut T;

  #[ensures(deref(result.as_ptr()) ==== value)]
  pub fn new(value: T) -> Cell<T>;

  #[ensures(deref(self.as_ptr()) ==== value)]
  pub fn set(&self, value: T);
}

#[extern_spec]
impl<T: Copy> Cell<T> {
  #[pure_unstable]
  #[ensures(result ==== deref(self.as_ptr()))]
  pub fn get(&self) -> T;
}
\end{rustblock}
\caption{Example of the capability and contract annotations on the \rust{Cell} type of the standard library. 
The figure contains two \rust{impl} blocks because in the second the \rust{T} parameter has an additional \rust{Copy} trait restriction; this is a design choice of the library API\@. 
All \rust{\#[capable(..)]} annotations are trusted by the verifier.
The method \rust{as_ptr} acts as a model of the memory location of the cell's content, so that other contracts can refer to it. 
}
\label{fig-04-approach-ext-cell-lib}
\end{figure*}

The program on the right of \Cref{fig-04-problems-example-refcell} contains two assertions, only one of which should verify.
Proving or disproving the assertions by using the library annotations provided in \figref{04-approach-ext-cell-lib} requires first reasoning about the value of \rust{x.as_ptr()}, which is used in the contracts of the \rust{Cell} library.
Since the \rust{Cell::as_ptr} method is marked as pure-memory, the call \rust{x.as_ptr()} is known to depend only on the values and corresponding addresses reachable from \rust{x}. 
These values and addresses are constant throughout the execution of the \rust{cell_client} function, because across each statement, a \rust{writeRef(x)} capability remains held by the client, implying an \rust{immutable(x)} capability that in turn guarantees immutability of \rust{x}.

Since across each statement \rust{writeRef(x)} implies \rust{readRef(*x)}, where \rust{*x} is of type \rust{Cell}, by the capability annotations of the library, we can then deduce \rust{local(x.as_ptr())} and \rust{noWriteRef(x.as_ptr())}.
These two capabilities guarantee immutability of the target of \rust{x.as_ptr()} across each statement of the function, except for \rust{use_cell(x)} because it is a call of a non-pure function.
This immutability property is sufficient to prove the last assertion in the program, as expected. 

Regarding the first assertion, checking \rust{a == b}, our technique does not generate any deduction that proves it, so the assertion is reported as potentially failing. 
This is the correct outcome because to falsify \rust{a == b} it would be enough to implement \rust{use_cell} with the single statement \rust{x.set(x.get() + 1)}.

\subsubsection{Usage of RefCell}

The program in \Cref{fig-04-problems-example-refcell} contains two assertions, the first of which can be verified by deducing an immutability property for the content of the \rust{RefCell} across the \rust{use_refcell(x)} call, the second by proving a non-aliasing property between the data contained in \rust{x} and \rust{y}.
To verify the absence of panics it would be necessary to also prove that the cell is not mutably borrowed when executing the \rust{borrow} method, because in that case the library would panic. 
The \rust{RefCell} type, in fact, contains two internal states mutable via shared references: one is used to store what is commonly called the content of the cell, while the second holds a reference counter, called borrow flag, that tracks how the content is borrowed. 
For simplicity, here we only present the main proof steps necessary to verify the two assertions, omitting the proof steps related to the borrow flag.

A simplified version of the annotated library is in \figref{04-approach-core-refcell-lib}.
Similarly to the annotations of \rust{Cell}, the method \rust{RefCell::as_ptr} is marked as pure-memory, thus the expression \rust{x.as_ptr()} can be deduced to be constant throughout the execution of \rust{refcell_client} because a \rust{readRef(x)} capability is available across each statement, and \readRef{} implies \immutable.

Initially, the content of the cell is copied into the variable \rust{before}, because \rust{*a} is desugared to \rust{*a.deref()} and the contract of the \rust{Ref::deref} function ensures that the result is a reference pointing to the memory location of \rust{a.as_ptr()}, which in turn is known to be equal to \rust{x.as_ref()} because of the postcondition of \rust{try_borrow}.
Across the initialization of \rust{before} and all the following statements the place \rust{a} remains \emph{immutably} available (\ie, available, but immutably borrowed), implying for each statement a \rust{readRef(a)} capability that, by the annotations of \rust{Ref}, in turn implies \rust{readRef(a.as_ptr())}. 
The immutability properties of the latter capability ensure that the content of the cell does not change across these statements, which in particular means that after \rust{use_refcell(x)}, the value of \rust{before} is still equal to the content of the cell.
Next, the \rust{*x.borrow()} expression returns a \rust{Ref} instance whose \rust{deref} method returns a reference to the content of the cell, which is then used to initialize the variable \rust{after}. 
Since the value of \rust{before} was already equal to the content of the cell, it thus follows that \rust{before} $=$ \rust{after}, as required to verify the first assertion.

For the second assertion, it is enough to notice that \rust{a} is available before the assertion, generating a \rust{writeRef(a)} capability that implies \rust{read(a.as_ref())} where \rust{a.as_ref()} $=$ \rust{x.as_ref()}. 
Also \rust{y} is available, generating a \rust{writeRef(y)} that, by the annotations of \rust{RefMut} and implication properties, leads to \rust{unique(y.as_ref())}. 
Because of the non-aliasing property between the implied \unique{} and \readC{} capabilities\footnote{An alternative is to use the non-aliasing property between the implied \immutable{} and \writeC capabilities.}, it follows that \rust{x.as_ref()} $\neq$ \rust{y.as_ref()}, which, considering the postcondition of \rust{RefMut::deref}, is then enough to prove that the second assertion always holds.

\begin{figure*}[htb]
\begin{rustblock}
#[capable(&self => readRef(self.as_ptr()))]
impl<'b, T> Ref<'b, T> {}

#[capable(&self => readRef(self.as_ptr()))]
#[capable(&mut self => writeRef(self.as_ptr()))]
impl<'b, T> RefMut<'b, T> {}

#[extern_spec]
impl<T> RefCell<T> {
  #[pure_memory]
  pub fn as_ptr(&self) -> *mut T;

  #[ensures(if let Ok(ref actual_ref) = result {
    actual_ref.as_ptr() == self.as_ptr() && ...
  } else { ... })]
  pub fn try_borrow<'b>(&'b self) -> Result<Ref<'b, T>, ...>;

  #[requires(...)]
  #[ensures(result.as_ptr() == self.as_ptr() && ...)]
  pub fn borrow<'b>(&'b self) -> Ref<'b, T>;
}

#[extern_spec]
impl<'b, T> Deref for Ref<'b, T> {
  #[pure_memory]
  #[ensures(result as *const _ == self.as_ptr())]
  fn deref<'a>(&'a self) -> &'a T;
}

#[extern_spec]
impl<'b, T> DerefMut for RefMut<'b, T> {
  #[ensures(result as *mut _ == self.data_ptr() && ...)]
  fn deref_mut<'a>(&'a mut self) -> &'a mut T;
}
\end{rustblock}
\caption{A simplified portion of the library specification of the \rust{RefCell} standard library module. Methods such as \rust{borrow_flag_ptr} are not part of the API of \rust{RefCell}, but clients of the library can model them by introducing new traits. The omitted code (\dots) corresponds to library properties that are not necessary to prove the examples in this paper.}
\label{fig-04-approach-core-refcell-lib}
\end{figure*}

\subsubsection{Usage of Arc}

The program in \figref{04-problems-example-arc} requires reasoning about the values and properties of the reference counter of the \rust{Arc} type, whose API is annotated in \figref{04-approach-example-arc-lib}.
Like with \rust{RefCell}, the type definition of \rust{Arc} actually contains (at least) two internal states mutable via shared references: one is used to store what is commonly called the content of the \rust{Arc}, while the second holds its reference counter, also called \emph{strong} counter\footnote{For simplicity, in this example we assume that the \emph{weak} counter is always zero. 
Reasoning about weak pointers in concurrent code is challenging, but we support that kind of reasoning in thread-unsafe libraries. For example, the \rust{Rc} library used in our evaluation.}.
Since the API does not expose the memory location of this counter, the specification models it with a pure method, marked as ghost to make sure that it can be called only from specifications\footnote{It does not need to be pure-memory, because moving an \rust{Arc} instance does not move its content or reference counter, so it is sound to abstract from the concrete memory.}.
The key capability annotation of the library is the following, expressing that if the value of the reference counter is $1$, then its value cannot be modified by other threads:
\begin{rustblock}
#[capable(&self if Arc::strong_count(self) == 1 => local(Arc::strong_ptr(self)))]
\end{rustblock}

Thanks to the conditionally \local{} capability of \rust{Arc} capability, in the first branch of \rust{arc_client}, the reference counter is known to be local.
Combined with the \noWriteRef{} capability of the \rust{Arc} library, it ensures that the counter remains exactly $1$ for the whole first branch, making it possible to verify the first assertion. 
Knowing that the counter is $1$, the second assertion can then be verified: before the assertion the explicit capability of \rust{x} implies \rust{writeRef(x.as_ptr())}, while the explicit capability of \rust{y} implies \rust{readRef(y.as_ptr())}.
The non-aliasing property between these two capabilities ensures that \rust{x.as_ptr()} $\neq$ \rust{y.as_ptr()}.
The assertion in the \rust{else} branch of the program checks a second time the same condition that was necessary to enter the branch. 
However, the evaluation of the condition of the \rust{if} and the evaluation of the last assertion are done at different program points. 
When the reference counter is not $1$, nothing in the library contracts guarantees its value remains the same. 
Other threads might drop all the existing clones of \rust{x}, suddenly bringing the reference counter to $1$ just before executing the last assertion. 
Without any information about the counter, a verifier will then conservatively report a verification error, as expected.

\begin{figure*}[htb]
\begin{rustblock}
#[extern_spec]
#[capable(&self => readRef(Arc::as_ptr(self)))]
#[capable(&self => noReadRef(Arc::strong_ptr(self)))]
#[capable(&self => noWriteRef(Arc::strong_ptr(self)))]
#[capable(&self if Arc::strong_count(self) == 1 => local(Arc::strong_ptr(self)))]
#[capable(&mut self if Arc::strong_count(self) == 1 => writeRef(Arc::as_ptr(self)))]
#[capable(&mut self if Arc::strong_count(self) == 1 => unique(Arc::strong_ptr(self)))]
impl<T> Arc<T> {
  #[pure]
  fn as_ptr(this: &Self) -> *const T;

  #[pure_unstable]
  #[ensures((result == 1) == (deref(Arc::strong_ptr(this)) == 1))]
  fn strong_count(this: &Self) -> usize;

  #[pure] #[ghost_fn]
  fn strong_ptr(this: &Self) -> *mut usize;
}
\end{rustblock}
\caption{A simplified portion of the specification of the \rust{Arc} type of the standard library. The \rust{strong_ptr} is not part of the API of \rust{Arc}, but it can be introduced using new traits. Here we show it as part of \rust{Arc} for simplicity. }
\label{fig-04-approach-example-arc-lib}
\end{figure*}

\begin{figure*}[htb]
  \begin{rustblock}
  fn use_refcell(x: &RefCell<i32>) { /* ??? */ }
  
  fn refcell_client(x: &RefCell<i32>, y: RefMut<i32>) {
    let Ok(a /*: Ref */) = x.try_borrow() else { return; };
    let before: i32 = *a;
    use_refcell(x);
    let after: i32 = *(x.borrow());
    assert!(before == after); // Succeeds
    assert!(x.as_ptr() as *const _ != y.deref() as *const _); // Succeeds
  }
  \end{rustblock}
  \caption{Example of a client of the \rust{RefCell} library. At runtime, both assertions never fail, so a complete verifier should be able to verify them. The first line in the implementation, \rust{let Ok(..) = ..}, makes the function return in case the \rust{try_borrow} method fails due to \rust{x} being already borrowed.}
  \label{fig-04-problems-example-refcell}
\end{figure*}

\section{Semantics \& Soundness}\labsec{04-soundness}

There are two aspects to the soundness of Mendel, the first is the soundness of the capabilities and Mendel's reasoning about them, and the second is the soundness of the trusted library specifications provided by Mendel. 
In this section, we address both aspects; we first present the intuitions behind Mendel's capabilities and then discuss a lightweight technique to prove that specific capabilities are sound.
Although a formal proof of soundness does not accompany Mendel, we present the intuitions justifying its underlying reasoning (and soundness) in this section. 
In this section, we sketch what a formalization of Mendel semantics would look like and the key theorems required to prove the soundness of our verifier. 

\paragraph{A primer on separation logic}
Separation logic is a formalism designed to reason about \emph{resources}, non-duplicable entities that can be shared or changed by different parts of a program. 
At its core is the concept of \emph{ownership}, typically of memory locations. 
When we say that $x$ owns the values $\bar{v}$ (also written $x \mapsto \bar{v}$), we mean that $x$ is the only name through which we can access the portion of memory described by $\bar{v}$.
Exclusive ownership enables mutation the memory locations of $x$. 
Ownership can be divided into arbitrary fractions (denoted $x \mapsto_q \bar{v}$), giving up the permission to mutate the memory in exchange. 

Occasionally, we must reason about \emph{shared mutable state}, whether for concurrency or to implement complex mutable data structures. 
Separation logic provides a tool for this: \emph{invariants}. 
An invariant is like a box containing an assertion describing a resource (such as $x$ is an even value). 
It can be temporarily opened for a single computation step, so long as we put the assertion back in afterward.
Between each opening, the resource inside may have changed; thus, we cannot be sure that two consecutive openings will yield the same result.
Invariants can be either \emph{global}, useful for concurrent reasoning, or \emph{thread-local}, which enables caches or other thread-local storage.

\subsection{Intuitive Semantics for Capabilities}

Mendel's capabilities have natural interpretations regarding separation logic: they are separation logic predicates, and their intuitions follow:
\begin{enumerate}
  \item \readC(p): Stores a fraction of the ownership for $p$ inside an invariant. 
  This allows the holder of this capability to open the invariant and read a value. 
  By quantifying existentially over the value of $p$ inside the invariant, we can ensure that two successive reads do not have the guarantee of observing the same value. 
  \item \writeC(p): Stores the \emph{whole} ownership for $p$ in an invariant.
  \item \immutable(p): Stores a fraction of the ownership for $p$ in a global an invariant.
  The difference with regards to \readC{}, is that it quantifies existentially over the value of $p$ \emph{outside} the invariant. 
  This ensures that each time the invariant is opened during a read, the same values \emph{must} be observed.
  \item \unique(p): Is an alias for ordinary ownership $p \mapsto \bar{v}$ of some values in memory. 
  \item \local(p): A \local capability ensures that the invariants of all \readC(p) capabilities is thread local. 
  \item \readRef(p): Contains the ownership predicate for an immutable borrow at $p$. 
  \item \writeRef(p): Contains the ownership predicate for a mutable borrow at $p$.
  \item \noReadRef(p): is equivalent to the negation of \readRef
  \item \noWriteRef(p): is equivalent to the negation of \writeRef
\end{enumerate}

\subsection{Soundness of Structural Properties}

Given appropriate definitions for each of the capabilities, the structural and implication properties can be modeled as ordinary separation logic implications. 
For example, weakening a \writeC(p) into a \readC(p) can easily be performed, since \writeC{} contains the whole ownership for $p$ in its invariant. 

This can also be extended to the non-aliasing properties: to verify that having both \writeC(p) and \immutable(p) it suffices to show that they would combine to more than whole ownership of $p$. 

\subsection{Soundness of Immutability Properties}

The immutability or framing properties used by Mendel are more challenging to specify. 
To prove these correct with the lowest required effort in a formalization like RustBelt, we propose to prove a series of lemmas corresponding to each individual immutability property. 
These lemmas would rely on being able to encode the equalities obtain by a given property as Rust assertions. 

Consider the property implied by \immutable(p), which ensures that across any statement, the value of $p$ is unchanged. 
Using the formalism of RustBelt~\cite{rustbelt}, the corresponding theorem for a program $F$ would look something like the following: given $\tctx \vdash F$ where $x \hasty \tau \in \tctx$, and $\immutable{x}$ then $\tctx \vdash \<let> x' = x \<in> F; \<assert>(x' = x)$ also types.
The proof for this would follow from the definition of \immutable{} which would allow us to conclude that the value of $x$ at the moment of the assertion must be the same as before the execution of $F$.

\subsection{Validating Library Specifications}
An essential task for any verifier is determining which values are unchanged or \emph{framed} by specific operations. 
In Mendel, this is one of the essential roles performed by the \readRef{} capability, which expresses the immutability of a reference to a memory location.
For example, consider the code on the left of \figref{04-soundness-rewriting}, which uses \rust{try_borrow} to perform a runtime aliasing check on a \rust{RefCell}, returning an \rust{a} of type \rust{Ref} if there are no outstanding \rust{RefMut} instances.
That \rust{Ref} instance carries with it a \readRef{} capability, which allows Mendel to conclude that the value of \rust{x} is not modified by the call to \rust{unknown}.

The capability annotations of \rust{RefCell} result from a careful study of the code and documentation of the \rust{RefCell} library, an ad-hoc process difficult to follow across Rust's many interiorly mutable libraries. 
We have developed a novel proof technique to help us specify the capabilities of Rust libraries.
The core idea is to rewrite a Rust program into a semantically equivalent one, making explicit the capabilities implicit in the source program. 

\begin{figure*}[htb]
\begin{minipage}{0.49\linewidth}
\begin{rustblock}
fn client(x: &RefCell<i32>) {
  if let Ok(a) = x.try_borrow() {
    // a is available from here...
    unknown(x);
    // ...to here
  }
}
\end{rustblock}
\end{minipage}
\hfill
\begin{minipage}{0.49\linewidth}
\begin{rustblock}
fn client(x: &RefCell<i32>) {
  if let Ok(a) = x.try_borrow() {
    let tmp: &i32 = a.deref();
    unknown(x);
    drop(tmp); // restores a
  }
}
\end{rustblock}
\end{minipage}
\caption{Example of a rewriting (original code on the left; resulting code on the right) that introduces a conversion method call, \rust{a.deref()}, using a place that was available across the \rust{unknown(x)} statement.}
\label{fig-04-soundness-rewriting}
\end{figure*}	

The code on the right of \figref{04-soundness-rewriting} performs this, introducing a local variable \rust{tmp} pointing to the content of \rust{x} and dropping \rust{tmp} after \rust{unknown(x)} so that it remains alive during the call.
This rewriting is semantically equivalent because \rust{deref} is a \emph{pure}, \emph{total} function that does not use synchronization primitives and thus has no observable side effects.
We call such methods \emph{conversion methods}.
At this point, we note that in Rust, it is UB's responsibility to have a shared reference whose target value is modified while the reference is alive.
Assuming that \rust{unknown} is sound, and given that \rust{tmp} is alive during the call of \rust{unknown}, its value (and thus \rust{x}'s) value cannot have been modified. 
This result holds in both the transformed and original programs as they are observationally equivalent, and \rust{unknown} cannot detect the rewriting.
This rewriting justifies the \readRef{} capability given to \rust{Ref} instances. 

Concretely performing in a tool the rewriting is not necessary in order to derive, \eg, the immutability properties of a \readRef{} capability across \rust{unknown(x)} in the example.
Instead, it is sufficient to identify that the rewriting is possible.

One assumption of our proof technique is to know what are the \emph{conversion methods} of a library API, which we define as the public safe methods whose only purpose is to convert between two types in a way that (i)~has no side effects, (ii)~always terminates, (iii)~never panics, and (iv)~does not use synchronization primitives. 
Intuitively, this means that usages of conversion methods cannot be detected at runtime by the library by means such as incrementing a counter each time the method is executed.
Examples of conversion methods in the standard library are the methods \rust{get_mut} (implemented for \rust{Mutex}, \rust{RwLock}, \rust{Cell}, \rust{RefCell}, \rust{OnceCell}, \rust{UnsafeCell}, \rust{SyncUnsafeCell}), \rust{Deref::deref} (implemented for \rust{MutexGuard}, \rust{RwLockReadGuard}, \rust{RwLockWriteGuard}, \rust{Ref}, \rust{RefMut}, \rust{Box}), and \rust{DerefMut::deref_mut} (implemented for \rust{MutexGuard}, \rust{RwLockWriteGuard}, \rust{RefMut}, \rust{Box})~\cite{rust-std-lib}.

Identifying the conversion methods of a library API is a manual process that requires reading the documentation and the source code of the library.
However, we believe that automatic identification of such methods is much easier than proving the immutability of a memory location handled via raw pointers in unsafe code, which is the alternative to our technique.

\section{Encoding}\labsec{04-encoding}

In this section, we present the main choices of our encoding of Rust programs and of the implicit capabilities to a first-order-logic subset of the Viper language~\cite{viper}. 
This Viper-based approach allows us to define our Rust verification technique independently from the lower-level techniques used for verification, such as symbolic execution or verification condition generation.

\subsection{Memory Model}

At a high level, our encoding uses a versioning technique to model the state of the memory at different program points in a control-flow graph.
Each program point is associated with an unconstrained memory.
Then, while encoding each statement in the control-flow graph, the encoding progressively introduces constraints between memory versions.
These constraints express the framing properties which hold across the execution of a statement.
Possible thread interferences are encoded as no-op statements, over which only the values other threads cannot modify are framed.
Additionally, the typing information generates non-aliasing constraints between values stored with the same memory version, similar to a separating conjunction in separation logic.

We represent the memory (both heap and stack) of a Rust program with a family of \emph{memory} total functions
$\mathcal{M}_{T}: \mathcal{A}_{T} \times \mathcal{V} \rightarrow \mathcal{S}_{T}$,
each of which, indexed by a type $T$, maps an address $\mathcal{A}_{T}$ and a memory version $\mathcal{V}$ to a \emph{memory snapshot} $\mathcal{S}_{T}$ (presented later).
These functions are defined using Viper \emph{domains} in the Viper language.

\subsection{Type Instances}

Rust type instances are modeled as \emph{memory snapshots}: a mathematical representation of the accessible values and memory locations without following raw pointers or entering \rust{UnsafeCell} types.
In our encoding, we model the memory snapshot of a variable \rust{x} as an algebraic data type (representing its reachable values and memory locations) with the following recursive definition:
\begin{itemize}
	\item The memory snapshot of a primitive type is its mathematical value.
	\item The memory snapshot of a raw pointer is the target's address, represented as an integer.
	\item The memory snapshot of a reference comprises both the address and the memory snapshot of its target.
	\item The memory snapshot of \rust{UnsafeCell} does \emph{not} contain any representation of its content. It is as if the \rust{UnsafeCell} were defined as an empty tuple.
	\item The memory snapshot of tuples is composed of the memory snapshot of its elements.
	\item The memory snapshot of structures is composed of the memory snapshot of its fields.
	\item The memory snapshot of an enumeration is composed of the discriminant represented as an integer and then, depending on the value of the discriminant, the memory snapshot of the fields of the corresponding variant.
\end{itemize}

Memory snapshots are necessary for modeling pure-memory Rust functions, encoded as a mathematical function whose arguments and return value are the memory snapshots of the corresponding Rust instances.
To model pure-value functions, we rely on a weaker representation of Rust instances that abstracts over memory addresses, called \emph{value snapshots}\footnote{For the readers who are already familiar with \citet{prusti-nfm}, the snapshots of Prusti correspond to the value snapshots of this encoding.}.
Since memory snapshots are always at least as descriptive as value snapshots, it is possible to convert instances of the former to the latter.

\subsection{Separation in First-Order Logic}

For each type, the approach described in \secref{04-approach} defines that a memory address parametrizes capabilities at a given program point.
This is sufficient in a separation-logic formalization, where non-aliasing properties can be defined as holding only between capabilities associated with separate portions of the memory.
However, for automation, we designed our encoding based on first-order logic, in which we model capabilities as uninterpreted boolean functions. 
In this setting, we need a new mechanism to encode whether two capabilities originate from different portions of the memory.

In our encoding, we model memory disjointness by introducing for all capabilities a new parameter that we call \emph{root place}, which represents with a syntactic Rust place the set of memory locations that can be reached and mutated using safe Rust code.
We define a (finite) set of root places for each program point so that different root places correspond to disjoint memory regions.
Thanks to the design of Rust, we can compute this set just by analyzing the borrow-checking information provided by the compiler.
The compiler already uses similar representations to ensure memory safety in safe Rust.

Given a type $T$, we define the capability functions $\ang{kind}_T(r, a, w)$, where $\ang{kind}$ is one of the capability kinds (\ie, \readRef, \writeRef, \etc), $w \in \mathcal{V}$ is a memory version that models the memory at a program point, $a \in \mathcal{A}_T$ is the memory address of the capability and $r \in \mathbb{N}$ identifies the root place at program point $w$ from which the capability originates.
In our encoding, each root place corresponds to an \emph{explicit} capability assumed to hold based on the compiler information.

With syntactic checks, we also determine which subset of the root places of a program point is not used by the statement that follows immediately after.
The \emph{unused root places} computed this way are useful in our encoding to model framing properties across the statement. 
They represent memory locations guaranteed to be unreachable (thus unaffected) by the statement.

\subsection{Properties of Capabilities}
For each Rust type $T$ used in a program, the properties of its capabilities are modeled with axioms.

\subsubsection{Implication Properties}
The implications represented by the arrow edges in \figref{04-approach-ext-capability-kinds} are modeled by axioms of the following shape, where $\ang{kind\(_\text{LHS}\)}$ and $\ang{kind\(_\text{RHS}\)}$ are the source and target capability kinds of the edge:
$$
\forall r \in \mathbb{N}, a \in \mathcal{A}_T, w \in \mathcal{V}. \;
\ang{kind\(_\text{LHS}\)}_T(r, a, w) \Rightarrow \ang{kind\(_\text{RHS}\)}_T(r, a, w)
$$

As an example, one of the implications for the \rust{i32} type is the following, expressing that for all roots, memory addresses, and versions, a \writeRef{} capability always implies \unique:
$$
\forall r, a, w. \; \texttt{writeRef}_\texttt{i32}(r, a, w) \Rightarrow \texttt{unique}_\texttt{i32}(r, a, w)
$$

The structural implications of \figref{04-approach-core-structural} are encoded by a slightly more generic axiom of the following shape, where $\ang{a\(_\text{RHS}\)}$ is the expression representing the address of a field instance or of the target of a reference:
$$
\forall r, a, w. \; \ang{kind\(_\text{LHS}\)}_T(r, a, w) \Rightarrow \ang{kind\(_\text{RHS}\)}_T(r, \ang{a\(_\text{RHS}\)}, w)
$$

\subsubsection{Non-Aliasing Properties}
Similarly to the implication properties, the non-aliasing represented by the red dashed edges in \figref{04-approach-ext-capability-kinds} are encoded as axioms of the following shape:
$$
\forall r_1, r_2, a, w. \; \ang{kind\(_\text{LHS}\)}_T(r_1, a, v) \land \ang{kind\(_\text{RHS}\)}_T(r_2, a, v)
\; \Rightarrow \; r_1 \neq r_2
$$

\subsubsection{Capability Annotations} The encoding of any capability annotation, conditional or not, is done by generating an axiom of the following shape.
In the axiom template, $\ang{cond}$ represents the runtime condition that guards the capability annotation (``true'' if there is none), and $\ang{a\(_\text{RHS}\)}$ is the encoding of the Rust expression identifying the target memory location of the capability.
$$
\forall r, a, w. \; \ang{kind\(_\text{LHS}\)}_T(r, a, w) \land \ang{cond}
\; \Rightarrow \; \ang{kind\(_\text{RHS}\)}_T(r, \ang{a\(_\text{RHS}\)}, w)
$$

\subsubsection{Immutability Properties}

The immutability properties constrain the memory versions of two program points separated by a statement, \eg, $w_1$ and $w_2$. 
In our encoding, we model each statement as having one extra memory version representing the transition between $w_1$ and $w_2$. 
The encoding assumes (based on the compiler's information) the explicit capabilities of the \emph{unused root places} of the statement. 
These follow all properties described above, plus immutability properties such as the following, which models the immutability of the \immutable{} capability.
$$
\forall r, a, w_1, w_2. \; \texttt{immutable}_T(r, a, \text{across}(w_1, w_2))
\; \Rightarrow \; \mathcal{M}_T(a, w_1) = \mathcal{M}_T(a, w_2)
$$

\section{Implementation and Evaluation}\labsec{04-evaluation}

\begin{table*}[tb]
	\caption{On the left: Description of the modeled unstable memory locations. %
	On the right: Description of the library annotations. ``LOC'' reports the number of lines of code needed to annotate the library (excluding empty lines, comments, imports and module declarations). ``Functions'' reports how many existing functions (including methods) were annotated as pure or not (``Imp.''), and how many ghost functions were necessary to annotate the library. ``Specifications'' reports the number of library capability annotations (``Cap.''), and the number of lines of code (``Contr.'') occupied by the contracts (\ie, pre- and postconditions, and purity attributes).}
	\labtab{04-evaluation-libs}
	\setlength{\tabcolsep}{4pt}
	\scalebox{0.8}{
	\begin{tabular}{p{3cm}p{4cm}}
		\toprule
		Library & Modeled unstable locations \\ %
		\midrule
		\rust{UnsafeCell}       & Content                         \\
		\rust{Rc}               & Content, strong and weak reference counters \\
		\rust{Arc}              & Content, strong reference counter \\ %
		\rust{Cell}             & Content                         \\
		\rust{RefCell}          & Content, borrow flag            \\
		\rust{AtomicI32}        & Content                         \\
		\rust{Mutex}            & Content, poison flag, lock flag \\
		\rust{RwLock}           & Content, poison flag            \\
		\midrule
		\rust{Box}              & Target                          \\
		\midrule
		\rust{Mutex} with invariant & - \\ %
		Verus-style cell            & - \\
		Verus-style pointer         & Target \\ %
		\bottomrule
	\end{tabular}
	}
	\hfill
	\scalebox{0.8}{
	\begin{tabular}{lrrrrrr}
		\toprule
		\multirowcell{2}[-0.4ex][l]{Type} & \multirowcell{2}[-0.4ex]{LOC} & \multicolumn{3}{c}{Functions} & \multicolumn{2}{c}{Specifications} \\
		\cmidrule(r){3-5} \cmidrule(l){6-7}
		{} & {} & \makecell{Imp.} & \makecell{Pure} & \makecell{Ghost} & \makecell{Cap.} & \makecell{Contr.} \\
		\midrule
		\rust{UnsafeCell}           & 16  & 3 & 1 & 0  & 1  & 6  \\
		\rust{Rc}                   & 89  & 4 & 4 & 2  & 10 & 33 \\
		\rust{Arc}                  & 51  & 2 & 3 & 1  & 6  & 18 \\
		\rust{Cell}                 & 28  & 3 & 2 & 0  & 4  & 8  \\
		\rust{RefCell}              & 127 & 6 & 1 & 10 & 8  & 52 \\
		\rust{Ref}                  & 34  & 1 & 1 & 2  & 3  & 9  \\
		\rust{RefMut}               & 45  & 2 & 1 & 2  & 5  & 11 \\
		\rust{AtomicI32}            & 23  & 2 & 0 & 1  & 3  & 5  \\
		\rust{Mutex}                & 103 & 4 & 1 & 8  & 9  & 40 \\
		\rust{MutexGuard}           & 48  & 3 & 0 & 2  & 6  & 12 \\
		\rust{RwLock}               & 76  & 5 & 1 & 5  & 4  & 35 \\
		\rust{RwLockReadGuard}      & 23  & 1 & 0 & 2  & 1  & 4  \\
		\rust{RwLockWriteGuard}     & 37  & 2 & 0 & 2  & 4  & 7  \\
		\midrule
		\rust{Box}                  & 40  & 3 & 1 & 1  & 3  & 10 \\
		\rust{Option}               & 33  & 3 & 2 & 0  & 0  & 22 \\
		\rust{Result}               & 37  & 4 & 2 & 0  & 0  & 25 \\
		\rust{ControlFlow}          & 9   & 0 & 2 & 0  & 0  & 4  \\
		\midrule
		\rust{Mutex} with inv. & 51  & 6 & 2 & 0  & 0  & 10 \\
		Verus-style cell     & 82  & 5 & 0 & 3  & 0  & 22 \\
		Verus-style pointer  & 74  & 4 & 1 & 3  & 2  & 16 \\
		\bottomrule
	\end{tabular}
	}
\end{table*}

\begin{table*}[tb]
	\caption{Description of the verified clients, each of which tests properties of the library type reported in the ``Used library type'' column. ``Lines of code'' reports the total number of lines of code of the program (excluding empty lines, comments, empty \rust{main} functions, imports and module declarations) and, among them, the lines of code used for contracts. ``Assertions'' reports the number of assertions to be verified in the program, classified by meaning: ``Expected'' are assertions that verify or that report a verification error as expected; ``Incompl.'' are assertions for which the verifier reports an error due to an incompleteness. Each assertion occupies 1 line of code, counted as part of the ``Total'' column. ``Time'' reports the average verification time with standard deviations, out of 10 runs, using the Viper backend based on verification condition generation.}
	\labtab{04-evaluation-clients-stats}
	\setlength{\tabcolsep}{4pt}
	\scalebox{0.8}{
	\begin{tabular}{llrrrrr}
		\toprule
		\multirowcell{2}[-0.4ex][l]{Client} & \multirowcell{2}[-0.4ex][l]{Used library type} & \multicolumn{2}{c}{Lines of code} & \multicolumn{2}{c}{Assertions} & \multirowcell{2}[-0.4ex]{Time (s)} \\
		\cmidrule(r){3-4} \cmidrule(l){5-6}
		{} & {} & \makecell{Total} & \makecell{Contracts} & \makecell{Expected} & \makecell{Incompl.} & {} \\
		\midrule
		\texttt{arc.rs}         & \rust{Arc<i32>}               & %
			66  & 0 & 27 & 0 & 10.0 $\pm$ 0.7 \\
		\texttt{arc\_rwlock.rs} & \rust{Arc<RwLock<Vec<i32>>>}  & %
			97  & 6 & 29 & 2 & 34.6 $\pm$ 0.8 \\
		\texttt{atomic.rs}      & \rust{AtomicI32}              & %
			35  & 0 & 9  & 2 &  5.6 $\pm$ 0.1 \\
		\texttt{cell.rs}        & \rust{Cell<i32>}              & %
			102 & 5 & 30 & 0 &  8.8 $\pm$ 0.2 \\
		\texttt{mutex.rs}      & \rust{Mutex<i32>}             & %
			47  & 0 & 18 & 0 & 12.7 $\pm$ 0.4 \\
		\texttt{rc.rs}          & \rust{Rc<i32>}                & %
			102 & 0 & 53 & 0 & 15.8 $\pm$ 0.9 \\
		\texttt{refcell.rs}     & \rust{RefCell<i32>}           & %
			71  & 6 & 25 & 0 & 13.5 $\pm$ 0.8 \\
		\texttt{unsafecell.rs}  & \rust{UnsafeCell<i32>}        & %
			35  & 7 & 7  & 0 &  4.8 $\pm$ 0.2 \\
		\midrule
		\texttt{box.rs}         & \rust{Box<i32>}               & %
			10  & 0 & 4  & 0 &  5.5 $\pm$ 0.1 \\
		\midrule
		\texttt{mutex\_inv.rs}  & \rust{Mutex<i32>} with invariant & %
			11  & 0 & 1  & 0 &  4.9 $\pm$ 0.3 \\
		\texttt{verus\_cell.rs} & Verus-style cell              & %
			9   & 0 & 4 & 0 &  6.2 $\pm$ 0.4 \\
		\texttt{verus\_ptr.rs}  & Verus-style pointer           & %
			34  & 7 & 10 & 0 & 10.3 $\pm$ 0.5 \\
		\bottomrule
	\end{tabular}
	}
\end{table*}

We implemented our verification technique in \emph{Mendel}: our new capability-based verification tool for Rust. 
As an engineering choice, we developed Mendel by reusing various components from the codebase of Prusti: parsing and type-checking of specifications, retrieval of borrow-checker information, and so on.
Our implementation removes support for many Rust features not directly relevant to our usage of capabilities such as closures, traits, loops, quantifiers, or iterators.
None of these are incompatible with our technique but would require additional engineering work to implement.
We then used Mendel to show that our capability annotations are \emph{useful} and, on the client side, \emph{lightweight}.
All measurements were averaged over 10 runs (after a JVM warm-up) on a laptop with a \texttt{i7-7700HQ} processor, 16 GB of RAM, and operating system Ubuntu 22.04.

\paragraph{Benchmark Origins}
Our benchmarks consist of interiorly mutable libraries and handwritten clients, which we gathered from several sources.
First, we annotated the APIs of popular types with and without interior mutability of the standard library using capability annotations, contracts, and helper ghost methods.
As a second source of libraries, we ported to our specification language the core of three libraries related to interior mutability taken from the test suites of Creusot and Verus: a \rust{Mutex} with a monitor invariant from the former; a cell-like and a pointer-like type from the latter.
The annotated libraries are described in \tabref{04-evaluation-libs}. 

For each library, we built several safe clients, each including many assertions to check the functional behavior of the API interactions.
These clients are simple compared to real-world code, but the goal is to test short sequences of API calls. 
Finally, we ran our verification tool on the clients, measuring their verification time. 
The client programs and their verification time are described in \tabref{04-evaluation-clients-stats}. 

\subsection{Results}

We now present the results of our evaluation for both libraries and their clients.

\subsubsection{Library Benchmarks}
In \tabref{04-evaluation-libs}, on the left, we listed the unstable memory locations we modeled for each library using our capability annotations. 
The simplest library types, such as \rust{UnsafeCell} or \rust{Cell}, only contain one memory location, while more complex types can have more. 
For example, we modeled three unstable memory locations in the \rust{Rc} and \rust{Mutex} libraries.
When annotating the \rust{Arc} library, we made one simplifying assumption: we assumed that the weak reference counter is always zero. 
This limitation does not apply to single-threaded libraries such as \rust{Rc}, for which we can annotate its strong and weak reference counter.
Regarding the types taken from the test suite of Creusot and Verus, we kept the existing assumptions: the type of a \rust{Mutex} with invariant assumes that no mutex is ever poisoned -- something that happens when a thread panics while holding a lock. 
This is not the case for our other annotations of the \rust{Mutex} and \rust{RwLock} types, for which we also model the panic flag. 
Regarding the Verus-style pointer type, its API is sound, assuming that its clients always respect the declared preconditions. 
This choice of Verus brings greater flexibility in the design of libraries. 
We retained this choice to evaluate our tool on these advanced specification cases.

In \tabref{04-evaluation-libs}, on the right, we present statistics regarding our annotations of the libraries.
We could reuse existing API methods for many library types by marking the methods as pure. 
For example, in the \rust{Rc} library, we marked the \rust{Rc::as_ptr} method as pure-value, \rust{Deref::deref} pure-memory, \rust{Rc::strong_count} and \rust{Rc::weak_count} as pure-unstable. 
However, the existing methods were not always sufficient to specify the library. 
In some cases, we needed to introduce new ghost methods to model specific type properties. 
Consider the \rust{Mutex} type: we added a \rust{data_ptr} method that returns the address at which the content of the mutex is stored.
We marked these additional methods as ghost, though the library implementation could provide most.

Most of our ghost methods were necessary to model aspects of the interiorly mutable values of libraries. 
For example, by exposing their address or by making it possible to refer to their value from specifications.
Among the types with the highest number of ghost methods, \rust{RefCell} uses them to expose the address and value of the contained data and of the borrowing flag, tracking the aliasing status of the type (non-borrowed, read-borrowed, or write-borrowed). 
In \rust{Mutex}, we used the ghost methods to model the address and value of the protected data and other internal flags (locking and poisoning). 
In \rust{RwLock}, we did the same as in \rust{Mutex}, but without modeling the locking flag.
In the Verus-style libraries, existing methods were already marked as ghosts.

\subsubsection{Client Benchmarks}
In \tabref{04-evaluation-clients-stats}, we report statistics regarding the clients in our evaluation, among which the used library type and the average verification time. 
Each function in the client programs makes a sequence of API calls, checking with many assertions or with a postcondition that the verifier can prove the expected functional behavior of the library.
We also used preconditions to check only a particular scenario in a few clients.
For example, in \texttt{refcell.rs}, we gave functions the precondition that their \rust{RefCell} argument is not read- nor write-borrowed.
In two cases, we also hit an incompleteness of our technique, described at the end of this section. 

The verification time we measured for these programs goes from 4.8 seconds for the most straightforward libraries to 34.6 seconds for the more complex ones.
The slowest client, \texttt{arc\_rwlock.rs}, requires the verifier to reason about nested library types: a \rust{Arc} containing a \rust{RwLock}, containing a \rust{Vec}.
The second slowest client, \texttt{refcell.rs}, uses the \rust{RefCell} library, which has the highest number of modeled unstable locations.
These measurements were all made using the Viper backend, which internally translates Viper programs to Boogie.
By profiling the verifier in a few cases, our preliminary results indicate that most of the verification time is spent generating and parsing Boogie code. 
This might result from our encoding technique that generates one axiom to model each capability property.
Future work might confirm these profiling results and explore more efficient engineering solutions.

\subsection{Discussion}

From the evaluation results, we can conclude that our specification technique works effectively on real-world libraries with interior mutability. 
It makes it possible to describe the properties decided by library developers and to verify the usage of these libraries using an automated verifier. In particular:
\begin{enumerate}
	\item The specification language is expressive: it is possible to explicitly declare the interior mutable properties of common standard library types.
	\item The technique is compositional: in the presence of nested types such as \rust{Arc<RwLock<Vec<i32>>>}, the capabilities automatically propagate properties across the type boundaries.
	\item The technique integrates with existing code: developers can reuse existing methods in the specifications, for example by modeling with the \rust{as_ptr} method the address of the content of the \rust{Arc} library. 
	Many types require no new methods to be added to the library.
	\item The technique is lightweight: in many cases, the client programs required no proof annotations, and the verification time was reasonable.
\end{enumerate}

Nevertheless, our verification technique has some limitations. 
Our technique uses capabilities to deduce a program's framing and non-aliasing properties. 
In certain situations, the capabilities we presented are not expressive enough to describe the content of a type with sufficient precision.
When this occurs, we must reason conservatively, assuming that the content might be mutated by any function call or aliased by any other type instance.
This leads to incompleteness when reasoning about the framing or non-aliasing properties of the program.
We encountered two cases during the evaluation where the verifier reported an incompleteness error.
For the \rust{Arc} type, we do not have a capability that precisely describes its content when (a)~the strong reference counter is not 1, or (b)~the strong reference counter is 1, the weak reference counter is 0, and the \rust{Arc} is immutably shared. 
For the \rust{AtomicI32} type, we do not have a capability that describes its content when the \rust{AtomicI32} instance is borrowed by \emph{local references}\footnote{That is, a reference that has not been passed to any function call.}.
Future work might overcome this by introducing new capabilities or combining our capability-based technique with other verification techniques for concurrent code.

\section{Related Work}\labsec{04-related-work}

\subsection{Rust Verifiers}

Prusti~\cite{prusti} is a deductive verifier for Rust that leverages Rust's type properties to build a memory-safety proof based on separation logic automatically and --- given user-written contracts --- verify the functional correctness of the program. Prusti does not support reasoning about interior mutability because it lacks the necessary expressivity and completeness, even though its verification technique is sound in the presence of libraries implemented with unsafe code.
Since both Mendel and Prusti are sound on their own but incomplete in different cases, combining both approaches would make it possible to reduce the incompleteness to only the cases where both techniques are simultaneously incomplete.

RustBelt~\cite{rustbelt} is a Coq formalization of Rust in which it is possible to model and verify the soundness of libraries. 
This work was later extended by RustHornBelt~\cite{rusthornbelt}, adding support for verifying functional correctness. 
While both works are based on the Iris~\cite{iris} framework and require manual proofs, our verification technique is automated and can be used by developers who do not have advanced knowledge of Coq or separation logic.
The language of RustBelt and RustHornBelt is more expressive than the capability specifications of our work but also more verbose. 
Proving the properties of our capabilities in Iris is an interesting research question for future work. We believe our capabilities are a useful user-readable abstraction that, in Iris, correspond to lemmas that derive the properties of a capability from the semantic invariant of a library type.

Creusot~\cite{creusot} is a deductive verifier that leverages Rust's type properties to verify functional properties. 
Creusot uses a technique based on \emph{prophecies} to encode Rust programs into first-order logic, using the Why3 language~\cite{why3}.
The technique that it uses is not based on notions of capabilities and only supports reasoning about interior mutability by wrapping the types behind an API with a monitor invariant.
Our technique makes it possible to reason more precisely about mutations to the content of types with interior mutability without needing to change existing methods' signatures or to wrap the types behind a new API\@.
Moreover, Creusot's technique does not support reasoning about memory addresses, while our work has first-class support for them.

Verus~\cite{verus} is another deductive verifier for Rust that, like Creusot, is based on a first-order logic encoding of Rust programs. 
One novelty of Verus is it use of the linearity and borrow checks of Rust to let the user manage separation-logic permissions by using regular (ghost) Rust variables.
In our view, Mendel's capability system can be seen as a generalization of the permissions tracked by Verus.
For example, Verus' \rust{PermData} type might be modeled in our technique as a structure providing a \unique{} capability for the target of the associated \rust{PPtr} type.
The approach of Verus requires defining custom libraries in which the types representing permissions show up explicitly as arguments or return types, while our technique makes it possible to annotate existing libraries without modifying their method signatures, only adding new methods. 
As a result, our technique requires fewer annotations on the client side and can be applied to existing Rust code.

Aeneas~\cite{aeneas} translates a subset of Rust into a pure lambda calculus, which can be verified using interactive theorem provers like F* or Lean.
Their technique explicitly does not support interior mutability or unsafe code. 
However, we believe library annotations like ours might be used to detect usages of interior mutability that could be translated into a pure functional language.

\subsection{Verification of Other Languages}

RefinedC~\cite{refinedc} verifies the functional correctness of C code by using a type system with ownership and refinement types, carefully designed so that the Coq proof of memory safety and functional correctness is automated and syntax-directed. Compared to our specification language, their type system is more complex and does not have a notion of immutability.

VCC~\cite{vcc} verifies low-level concurrent C code annotated with global invariants~\cite{admissible-invariants}. Their invariants typically require each shared object to keep track of its referencing objects using a set of back-pointers. This technique could be ported to Rust by modeling a ghost set of back-pointers for each object.
However, cyclic data structures are unidiomatic in Rust, and manually updating the set of back-pointers is verbose. 
Our technique requires neither of the two. Still, our first-order logic encoding based on memory versions is inspired by their encoding to Boogie~\cite{boogie}.

Pony~\cite{pony} is a programming language that ensures data-race freedom of concurrent actor-based code using a strong type system with capabilities, among which deny and unique properties. 
The rich expressivity of Pony types inspired part of our work on the capabilities of Rust libraries.

\section{Conclusion}\labsec{04-conclusion}

We have presented a new technique to specify the capabilities of Rust libraries implemented with unsafe code and to verify the functional correctness of some of their safe clients.
Our approach enables developers to explicitly declare the intended aliasing and mutability properties of \emph{existing} libraries, making it easier for automated verifiers and other human developers to reason about their usage.
An open-source implementation of our verification technique and our evaluation data is available on GitHub~\cite{mendel-repo}.

\bibliography{bibliography}

\end{document}